\documentclass[conference]{IEEEtran}
\usepackage{listings}

\ifCLASSINFOpdf
\else
\fi

\usepackage{graphicx}
\usepackage{makecell}

\begin{document}
%
\title{Modelling the earth’s geomagnetic environment on Cray machines using PETSc and SLEPc}

\author{\IEEEauthorblockN{Nick Brown}
\IEEEauthorblockA{EPCC\\University of Edinburgh\\
Bayes Building, Edinburgh}\\
\IEEEauthorblockN{Susan Macmillan}
\IEEEauthorblockA{British Geological Survey\\The Lyell Centre\\Edinburgh}
\and
\IEEEauthorblockN{Brian Bainbridge}
\IEEEauthorblockA{British Geological Survey\\The Lyell Centre\\Edinburgh}\\
\IEEEauthorblockN{William Brown}
\IEEEauthorblockA{British Geological Survey\\The Lyell Centre\\Edinburgh}
\and
\IEEEauthorblockN{Ciar\'an Beggan}
\IEEEauthorblockA{British Geological Survey\\The Lyell Centre\\Edinburgh}\\
\IEEEauthorblockN{Brian Hamilton}
\IEEEauthorblockA{British Geological Survey\\The Lyell Centre\\Edinburgh}

}


%


\maketitle


\begin{abstract}
The British Geological Survey's global geomagnetic model, Model of the Earth's Magnetic Environment (MEME), is an important tool for calculating the earth's magnetic field, which is continually in flux. Whilst the ability to collect data from ground based observation sites and satellites has grown rapidly, the memory bound nature of the code has proved a significant limitation in modelling problem sizes required by modern science. In this paper we describe work done replacing the bespoke, sequential, eigen-solver with that of the SLEPc package for solving the system of normal equations. This work had a dual purpose, to break through the memory limit of the code, and thus support the modelling of much larger systems, by supporting execution on distributed machines, and to improve performance. But when adopting SLEPc it was not just the solving of the normal equations, but also fundamentally how we build and distribute the data structures. We describe an approach for building symmetric matrices in a way that provides good load balance and avoids the need for close co-ordination between the processes or replication of work. We also study the memory bound nature of the code from an irregular memory accesses perspective and combine detailed profiling with software cache prefetching to significantly optimise this. Performance and scaling characteristics are explored on ARCHER, a Cray XC30, where we achieved a speed up for the solver of 294 times by replacing the model's bespoke approach with SLEPc. This work also provided the ability to model much larger system sizes, up to 100,000 model coefficients, which is also demonstrated. Some of the challenges of modelling systems of this large scale are explored, and mitigations including hybrid MPI+OpenMP along with the use of iterative solvers are also considered. The result of this work is a modern MEME model that is not only capable of simulating problem sizes demanded by state of the art geomagnetism but also acts as further evidence to the utility of the SLEPc libary.
\end{abstract}


%
\IEEEpeerreviewmaketitle

\section{Introduction}

The British Geological Survey (BGS) global geomagnetic model inversion code, known as the Model of the Earth's Magnetic Environment (MEME) \cite{meme}, is used to produce various models of the earth's magnetic field. Written in Fortran 90, it is essentially a mathematical model of the earth's magnetic field in its average non-disturbed state. The input consists of millions of data points collected from satellite and ground observatories on or above the surface of the earth, which are used to identify the major sources of the magnetic field which include the core, crust, ionosphere, and magnetosphere. The magnetic field is then solved for the Gauss coefficients, which describe the magnetic field as weighting factors for spherical harmonic functions of a certain degree and order such as spatial wavelength. Additionally, the Gauss coefficients have a temporal dependence requiring the solution of weights for a sixth-order B-spline function. The output is a set of, as it currently stands, around 10,000 coefficients describing the spatial and temporal variation of the magnetic field from the core to near-earth orbit over a period of around 15 years. This allows a compact representation of the magnetic field.

From this single geomagnetic code, the community produces a number of models each year both for research and for non-scientific users. One such model is the International Geomagnetic Reference Field (IGRF) \cite{igrf} which is widely accepted as a standard, low spatial, resolution model of the earth’s magnetic field. The IGRF model alone has numerous users from solar-terrestrial physicists, who use it for their magnetic coordinate systems, to geomagneticists studying the history of the earth's magnetic field over thousands of years in order to understand the underlying physics generating the field. Another widely-used model derived from the MEME code is the World Magnetic Model (WMM) \cite{wmm} which is used in civilian and military navigation and positioning systems, including the vast majority of mobile phones. Other models are produced annually from the parent MEME model, for instance providing capabilities such as detailed navigation where very accurate values of declination are required, or to look at rapid time variations of the magnetic field for scientific study.

The MEME model has been around for a number of years and, although partially parallelised, has critical parts which currently run in a serial fashion, in particular the solution of the normal equations. This solution of the normal equations follows a bespoke eigen-solver approach and now the community wish to study systems much larger than the 10,000 coefficients which the current code is limited to. The serial portion of the code places significant limitations on scaling the spatial and temporal resolution of the model. The scientific impact is significant, as increased resolution means reduced uncertainties and an improvement in the predicted confidence levels of the modelled field. 


In this paper we describe work done modernising the MEME model using the PETSc \cite{petsc} and SLEPc \cite{slepc} library to parallelise the solving of the normal equations. In order to support the distribution of crucial data-structures, which is required for the scaling of the number of coefficients, new algorithms and approaches to building distributed symmetric matrices have been developed. A fundamental driver to this work was the memory bound nature of the previous model, both in terms of fitting into available memory, and also irregular memory accesses which limit cache effectively. In short, the contributions of this paper are

\begin{itemize}
\item The use of SLEPc for very large problem sizes where the matrix is over 100,000 by 100,000 elements and we are looking to find the large majority of Eigen values
\item A novel approach to building a symmetric matrix in a distributed fashion which requires minimal co-ordination and results in good load balance and avoids duplication of computation
\item How the role of software prefetching can assist codes that rely on irregular accesses, where the hardware prefetchers and cache organisation does a poor job
\end{itemize}

The layout of this paper is as follows, in Section \ref{sec:bg} we describe the background of the code in more detail and related work including technologies that have been used. In Section \ref{sec:matrixandrhs} we discuss our novel approach to distributing the building of the symmetric matrix across processes in a manner that provides good load balance and avoid duplicate calculation, along with concerns around irregular memory access and how we use software prefetching to mitigate these. In Section \ref{sec:slepcwork} we briefly describe the work done integrating the PETSc and SLEPc toolkits into the code, and then explore the performance of these in contrast to the previous model in Section \ref{sec:performance}. Challenges faced and solutions found to scaling the problem size to large numbers of coefficients are described in Section \ref{sec:largecoefs} before drawing conclusions and discussing further work in Section \ref{sec:conclusions}


\section{Background and related work}
\label{sec:bg}
\subsection{Previous model performance and scaling}

Before solving the normal equations, these need to be built. This boils down to building a matrix and Right Hand Side (RHS), and previously, a partial parallelisation of this code was undertaken with MPI. This concentrated on parallelising the building of these data structures, but with all other aspects, such as the solving of the equations, remaining serial. This previous parallelisation works by allocating the entire normal equation data structures, i.e. the entire matrix and RHS, on every process and decomposing on the input data. Contributions from each piece of input data are additive, and as such each process calculates the contributions for its subset of data across the entire matrix and RHS, before all individual processes' data is reduced (summed together) at rank 0.

\begin{figure}
	\begin{center}
		\includegraphics[scale=0.50]{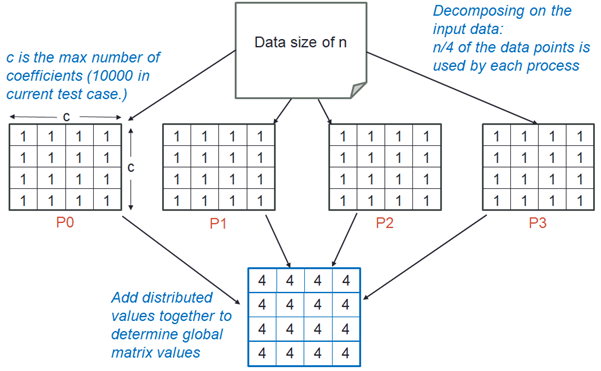}
	\end{center}
	\caption{Domain decomposition of the previous model}
	\label{fig-existing-decomp}
\end{figure}

Figure \ref{fig-existing-decomp} illustrates the previous model's parallelisation in more detail, where an input data size of \emph{n} is decomposed across the processes and it can be seen that the entirety of the matrix data structure is built locally for the processes' input data. Whilst the RHS is also build, this is fairly trivial, and it is the matrix of normal equations which presents the major difficulty here. Because the entire global matrix has to be held on each process, the memory limit of the machine is quickly reached. In Figure \ref{fig-existing-decomp} this is illustrated by \emph{c}, which is the number of model coefficients and determines the size of the matrix (\emph{c} by \emph{c}). The state of the art coefficient size is 10,000, which results in a matrix of size 10000 by 10000. Bearing in mind values are double precision, this requires 800MB of memory per process. The community would like to extend the model to much larger system sizes, but the amount of memory required increases as a square of the number of model coefficients, for instance, increasing to 20,000 coefficients would require 3.2GB memory per process. This is especially important because scientists are currently forced to throw away significant amounts of their input data, one such example being the EASA SWARM satellites where geomagneticists can only currently use around five percent of the data points collected due to the memory limitations of the model. They estimate that if the model could support 100,000 model coefficients, which is ten times the number of coefficients that it can currently handle and a matrix of normal equations a hundred times larger, they could take advantage of all the data collected by modern geomagnetic instruments. In addition to the memory limits, there is also a significant work imbalance between the processes in building the normal equations and from experimentation with the previous model we found a 37\% difference in the run time between the slowest and fastest process.

\begin{figure}
	\begin{center}
		\includegraphics[scale=0.50]{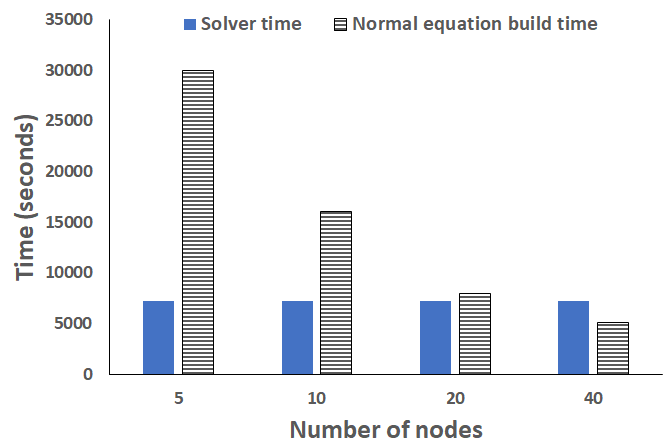}
	\end{center}
	\caption{Scaling characteristics of the previous model on ARCHER, a Cray XC30, with a problem size of 10,000 coefficients and 4.3 million input data points}
	\label{fig-existing-scaling}
\end{figure}

Figure \ref{fig-existing-scaling} illustrates performance and scaling of the previous model on ARCHER, a Cray XC30, for an experiment of 10,000 coefficients and 4.6 million data points. It can be seen that the solver time is constant, 7200 seconds, irrespective of parallelism and this is due to its sequential nature. The building of the normal equations does scale as parallelism increases, but at 516 cores still takes a significant amount of time. In addition to the load imbalance, there are other issues causing overhead with the previous model's normal equation building, one such example being the use of MPI P2P communications and then manually summing up values on the root, rather than an MPI reduce call, for the normal equation matrix reduction between processes. An eigen-solve approach is used to solve the normal equations, where a direct solver finds all the eigenvalues and eigenvectors, before applying these to the normal equations RHS to generate the solution. The previous model uses a bespoke, serial, Givens reduction which is highly stable and reliable \cite{givens}, but known to exhibit significant computational overhead \cite{givens-slow}.

\subsection{PETSc and SLEPc}
\label{sec:petscslepc}
The Portable Extensible Toolkit for Scientific Computation (PETSc) is a is a suite of data structures and routines developed for the parallel solution of scientific applications modelled by partial differential equations. Not only does PETSc ship with a variety of highly optimised pre-conditioners and solvers, there is also extensive support for flexibly selecting different modes of execution that include running serially, running over distributed memory machines (using MPI) and even GPUs. 

Whilst PETSc does not come with eigen-solving functionality as part of the main distribution, the library itself has been designed so that it can be used as building blocks by other libraries. The Scalable Library for Eigenvalue Problem Computations (SLEPc) sits on-top of PETSc, providing eigen-solver capability and relying on the PETSc eco-system for parallelisation, utility functionality and general program flow. A major benefit of using SLEPc is that the code is still written in the PETSc style, and the only difference needed is that the solver created is an eigen-solver in the SLEPc library rather than PETSc iterative solver. This is important, because not only does it mean that those familiar with the popular PETSc package can easily understand the code, but also it should be fairly trivial, from a code perspective, to swap out the eigen-solver and replace it with a PETSc iterative solver. The SLEPc library ships with a number of eigen-solvers including Krylov-Schur, Arnoldi, Lanczos, GD and Lapack, any of which can be selected by the user.

Whilst the PETSc library is provided as a module as part of the Cray Programming Environment, SLEPc is not and needs to be build specifically. However, it is trivial to build both of these libraries for the Cray eco-system and actually for the experiments detailed in Section \ref{sec:performance}, we built both PETSc and SLEPc specifically. This was because a newer version of PETSc contained a number of important bug fixes, and this had not yet been released via Cray's official programming environment. Regardless, this still takes advantage of the Cray scientific libraries and as such we saw a negligible difference in performance. 

There are a number of alternative libraries for solving the eigen value problem, one such example being Eigen \cite{eigen} which is a C++ template library for linear algebra and supports, amongst many other things, computing eigen value problems. Whilst this popular library is widely used and has excellent performance \cite{eigen-usage}, there are two major drawbacks in the context of the MEME model. Firstly it is not parallelised beyond threading and as-such unable to take advantage of distributed memory machines which is a major aspect here to increasing the number of model coefficients and scientific utility of the model. Secondly there are no bindings available for Fortran and it is not realistic to port the whole of the MEME model over from Fortran 90 to C++.

Another alternative approach for finding the eigenpairs (eigenvalues and eigenvectors) is the Eigenvalue SoLvers for Petaflop-Applications (ELPA) \cite{elpa} library. This package provides support for finding Eigenvalues and eigenvectors of large symmetric, Hermitian, matrices. Building on ScaLAPACK, ELPA provides its own highly efficient parallel implementations and has been shown to scale well to over 290,000 cores on a BlueGene/P \cite{elpa}. The library has bindings for Fortran, but a disadvantage is that the matrix must be Hermitian and some future problems that the MEME model will be used to target do not share this property.

We adopted the PETSc/SLEPc approach because PETSc is already installed on Cray machines, fairly ubiquitous in HPC, known to work well and the flexibility provided by these libraries is important. SLEPc has also been demonstrated to work well in a variety of application areas including engineering \cite{slepc-cfd}, materials science \cite{slepc-material-science}, structural analysis \cite{slepc-structural}, earth sciences \cite{slepc-earth} and geomagnetism \cite{slepc-geomag}. This last use-case is important because it is the same domain as the MEME model and as such SLEPc has already demonstrated benefit in the domain of geomagnetism.

\section{Distributed building of the normal equations}
\label{sec:matrixandrhs}
\begin{figure}
	\begin{center}
		\includegraphics[scale=0.50]{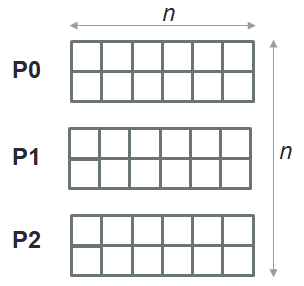}
	\end{center}
	\caption{Row based decomposition of the matrix for n model coefficients}
	\label{fig-matrix-decomp}
\end{figure}

Before solving the normal equations these must first be built. This involves calling user provided procedures which generate values, specific to the problem the user is looking to model, that ultimately populate the matrix and RHS data structures. The previous model placed the entirety of the matrix and RHS data structures on each process, decomposing via the input data instead. As described in Section \ref{sec:bg}, this significantly limits the problem size that can be modelled because, as one scales the number of model coefficients, the size of the matrix and by extension memory requirements, increase significantly. PETSc assumes a matrix decomposition based on rows, where a subset of rows of the matrix and elements of the RHS are held on each process on each process. Therefore to take advantage of PETSc, and model larger problem sizes, the approach to building the normal equations had to change. In fact the building of the RHS is fairly trivial, so for brevity in this section we focus on the building of the matrix which is far more challenging and the PETSc decomposition of the matrix, by row, is illustrated in Figure \ref{fig-matrix-decomp} for \emph{n} model coefficients. 

\begin{figure}
	\begin{center}
		\includegraphics[scale=0.40]{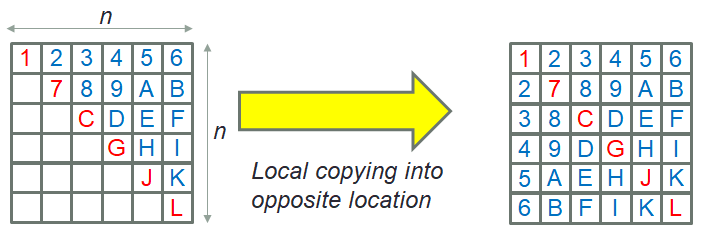}
	\end{center}
	\caption{The previous model just computes the diagonal and upper elements, then copies the upper elements into their corresponding lower position}
	\label{fig-matrix-copy}
\end{figure}

The challenge with this approach is that the matrix is in-fact symmetrical, so only the diagonal and one half of the matrix actually to be computed. It is trivial in the sequential case, where no matrix decomposition takes place, to take advantage of this property. A process simply computes the values for the diagonal and upper half of the matrix before copying the upper half to the lower half as illustrated by Figure \ref{fig-matrix-copy}. In previous version of the model each local process only calculates cell values for their  diagonal and upper parts  of the matrix before communicating these to the root (rank 0). Once the root has reduced, summed up, the values at each cell of the diagonal and upper half of the matrix from each process, it copies the upper values to their corresponding lower value location in the matrix. Because the building of the matrix and RHS is so expensive, see Figure \ref{fig-existing-scaling}, being able to limit the amount of data explicitly computed is an important saving.

\begin{figure}
	\begin{center}
		\includegraphics[scale=0.40]{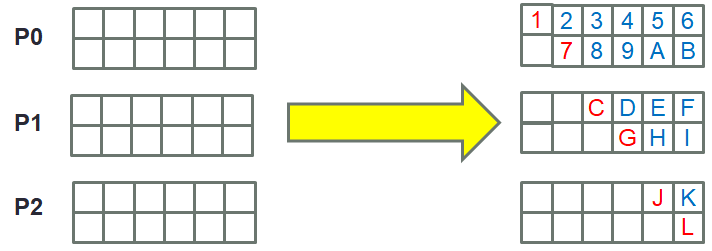}
	\end{center}
	\caption{Naive approach of only calculating diagonal and upper matrix points held locally}
	\label{fig-naive-local-copy}
\end{figure}

When it comes to building the matrix in parallel, the existing approach of only building the upper and diagonal parts of the matrix held on a process as per Figure \ref{fig-naive-local-copy}, and then communicating the upper values to the corresponding process, is not ideal. The reason for this can be seen in Figure \ref{fig-naive-local-copy}, where there would be a significant amount of load imbalance, for instance in this example, process zero must calculate 13 points whereas process two only 3 points. Because of the intensive nature of building the matrix this situation will result in process 2 spending much of its time idle. Another approach could be to build all points in the matrix held locally, irrespective of whether they are in the upper, diagonal or lower parts of the matrix. This avoids the load imbalance but because of the symmetry of the matrix, this approach involves redundant computation, which is especially inefficient due to the intensive nature of building the matrix. In contrast to the previous model, significant amounts of additional computation would be required, all of it redundant.

We therefore decided that a different approach was required which would naturally balance the load between processes and avoid redundant computation when building the matrix. Instead of simply calculating the diagonal and upper parts of the matrix held on a process, we developed an approach where processes would build they diagonal elements and specific upper and lower parts of their matrix, but crucially without any global duplication of cell calculation when it comes to symmetry. For instance, if one process is building the values for a lower cell then there is the guarantee that the corresponding upper cell value will not be built by the other process that holds it. For efficiency one of they challenges here was to do this in a way that would minimise the need for overarching co-ordination between processes deciding who builds what.

\begin{figure}
	\begin{center}
		\includegraphics[scale=0.70]{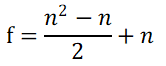}
	\end{center}
	\caption{The global number of cells that been to be explicitly calculated, where n is the matrix size in one dimension (the number of coefficients)}
	\label{fig-global-num-cells}
\end{figure}

In our approach we first calculate \emph{f}, the global number of cells that must be explicitly calculated based on a matrix of \emph{n} by \emph{n}. This corresponds to a problem size of \emph{n} coefficients and Figure \ref{fig-global-num-cells} illustrates the formula used. Using the example of Figure \ref{fig-naive-local-copy}, the matrix size (\emph{n} by \emph{n}) is 36 but \emph{f} is 21, which means that whilst there are 36 cells, only 21 values need to be explicitly calculated and any more would be duplicating work. 

\begin{figure}
	\begin{center}
		\includegraphics[scale=0.50]{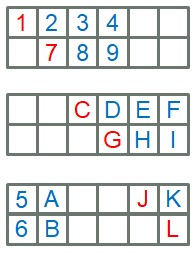}
	\end{center}
	\caption{Balanced building of the matrix}
	\label{fig-balanced-matrix-build}
\end{figure}

For each row of the matrix we then calculate \emph{r} which is \emph{f} divided by \emph{n} and call this the base number of points per row that needs to be calculated. Using the example of Figure \ref{fig-global-num-cells}, \emph{r} is 3.5 . Once this is calculated, for each row held locally by the specific process, this process starts at the diagonal element and calculates values for \emph{r} grid cells, which might wrap around from the upper to lower part of the matrix, as illustrated in Figure \ref{fig-balanced-matrix-build}. If \emph{r} is a whole number, which effectively means that \emph{n} is an odd number, then nothing further is required. 

However if \emph{r} is a fraction, as is the case in Figure \ref{fig-balanced-matrix-build}, an extra stage is needed. Firstly, for each row held locally the process will alternate \emph{r} between the ceiling of \emph{r} and the floor of \emph{r}. The first global row starts with the ceiling of \emph{r} and each process determines whether to start with ceiling or floor in reference to where its first row is in relation to that. Furthermore, if the number of rows divided by two (e.g. \emph{n} over 2) is even, then the ordering of ceiling and floor must be swapped for the second half of the matrix. This approach is illustrated in Figure \ref{fig-balanced-matrix-build}, where the elements per row alternates between ceiling and floor because \emph{r} is a fraction (3.5). Because \emph{r} divided by two is not even (6 divided by 2 equals 3), then no swapping of the order half way through is required. 

This algorithm provides the ability to build the matrix in a way that requires minimal coordination between the processes determining which cells they explicitly calculate, the avoidance of any replicated computation and reasonable load balance. Once local values have been calculated, as per Figure \ref{fig-balanced-matrix-build}, the next step is to then send upper or lower cell values that the process has computed to the other, corresponding, process which, due to matrix symmetry, requires that value in the opposite side of the matrix. Each process calculates both the number of values to send to, and the number of values to receive from, every other process. Based on these numbers, send and receive buffers are pre-allocated, and cell data, once calculated, is packed into the appropriate send buffer. Once this has been completed a single non-blocking MPI send and single non-blocking MPI receive are issued to every other processes if required (i.e. the number of cells to communicate is greater than zero). Therefore, for performance, at most there is a single large message sent and another single large message received, by each pair of processes.

\begin{figure}
	\begin{center}
		\includegraphics[scale=0.50]{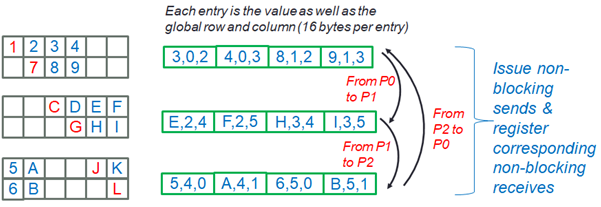}
	\end{center}
	\caption{Sending cell data to corresponding process}
	\label{fig-sending-matrix-cell-data}
\end{figure}

The packing of the send buffer and sending data is illustrated in Figure \ref{fig-sending-matrix-cell-data}. In addition to the double precision cell value we also send the zero indexed global row and column of that piece of cell, both of which are integers. Therefore the communication of a single cell requires 16 bytes, 8 bytes for the double precision cell value and two 4 byte integers holding the row and column location. This additional global coordinate is associated with each data value to ensure the correct mapping of values to cells on the receiving process. It also means that the job of the receiver, unpacking the receive buffer, is trivial because it just needs to translate the global coordinates to local coordinates and swap the row and columns round to identify the correct local location to write the cell value into. This addition of the global cell and row index, whilst it doubles the size of the message, makes the code significantly simpler as we do not need to worry about the explicit ordering of the buffer when packing values as the receiver obtains the location from the message itself rather than the cell's location in the message. As, generally speaking, for small and medium sized messages the message size in MPI does not make a huge overall difference to the communication performance \cite{mpi-message-size}, we felt that this was a reasonable design decision.

\begin{figure}
	\begin{center}
		\includegraphics[scale=0.50]{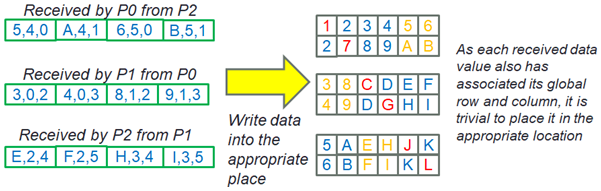}
	\end{center}
	\caption{Receiving values and copying them from the receive buffer into the appropriate matrix cell location}
	\label{fig-recv-matrix-cell-data}
\end{figure}

The MPI send and receive calls issued by the code are non-blocking, and whilst the communication is on-going each process performs local copies of cell values between corresponding upper and lower parts of the matrix that are held on the same process. For instance, in the example of Figure \ref{fig-global-num-cells}, the value 2 will be copied on process 0 from the 1st column of row zero to the zeroth column of row one. Once local copying is complete MPI \emph{waitany} calls are issued for the non-blocking communications and as non-blocking receives complete the received values are transferred from the receive buffers into the appropriate local matrix cell. This copying of both local and received data is illustrated in Figure \ref{fig-recv-matrix-cell-data} and once this has completed the matrix is fully built and the normal equations are ready to be solved.

\subsection{Improving memory locality}

In the original code over 35\% of the run time when building the normal equations was in a procedure which applies a batch of locally computed normal equations to each elements in the matrix according to some weight. When we profiled the code we found that over three quarters of the time spent in this procedure was due to the CPU stalled waiting for data from memory and this is due to irregular memory accesses. This code is sketched in Listing \ref{lst:irregularaccess}, and it can be seen that both for the
\emph{matrix} array which is being written to and the \emph{equations} array that is being read from, the index being used to access these data structures is indirect. In the case of the \emph{matrix} array the first index is based on the \emph{i}th value held in the \emph{dataloc} array and the for \emph{equations} array the index is based on the \emph{i}th value held in the \emph{inputdata} array. This is problematic when is comes to caching because, not only are these memory accesses irregular due to values in the \emph{dataloc} and \emph{inputdata} arrays changing significantly from one location to the next, which negates taking advantage of bringing in entire cache lines of data into the cache, but also unpredictable which negates use of the CPU's hardware prefetching.

\begin{lstlisting}[frame=lines,caption={Illustration of codes irregular memory access},label={lst:irregularaccess},numbers=left]
do j=1, n
    do i=1, n
        matrix(dataloc(i), j)=equations(inputdata(i,j)) + ....
    end do
end do
\end{lstlisting}

This is further illustrated in Figure \ref{fig-poor-counters-irregular}, and the table illustrates statistics gathered for the code from the CPU's hardware counters. The first six rows represent raw hardware counters, and the last three rows metrics derived from these. The number of cycles is divided by the number of instructions issued to produce the cycles per instruction (CPI) and vice-versa the number of instructions issued is divided by the number of cycles to determine the number of instructions per cycle. The \emph{total number of resource stalls} illustrates the number of cycles that the front-end of the CPU was stalled, due to running out of resources, and the \emph{Resource stalls (store buffer)} is the number of cycles the front-end of the CPU was stalled due to running out of store buffer entries. The \emph{cycles no instructions are issued} represents the number of cycles where no instructions are issued to execution units, for whatever reason. This is used, in conjunction with the number of cycles, to determine the total percentage of cycles that the CPU is stalled. It can be seen in Figure \ref{fig-poor-counters-irregular} that the CPU is idle for many cycles, there are many front-end resource stalls due to the CPU running out of store buffers and on average it takes 2.26 cycles to execute one instruction, or on average every cycle we complete 0.44 of an instruction. The Ivy Bridge micro-architecture contains five execution units, so there is a theoretical peak of 5 instructions per cycle or 0.2 CPI. Reasons for not reaching this theoretical peak are not just based on memory limitations, for instance many similar instructions contenting for specific execution units can have an impact, but still from the other metrics we can see that the CPU is stalled for a significant amount of time and this goes partly to explaining why we are currently such a long way off the theoretical peak.

\begin{figure}
\begin{center}
\begin{tabular}{ | c | c | }
\hline
Counter description & Value \\ \hline
Number of cycles & 69,109,605,287 \\
Number of instructions issued & 30,451,871,184\\
Total resource stalls & 60,734,999,355\\
Resource stalls (store buffer) & 60,042,957,250\\
Cycles no instructions are issued & 54,243,641,893\\
L1 cache hits & 5,613,708,007\\
\hline
Instructions per cycle  & 0.44\\
Cycles per instruction & 2.26 \\
Total \% cycles stalled & 78 \\
\hline
\end{tabular}
\end{center}
\caption{Hardware counter values for irregular memory access}
\label{fig-poor-counters-irregular}
\end{figure}

However, when it comes to improving this situation the options are not necessarily simple. This is largely in part due to the Out-of-Order (OoO) execution nature of modern CPUs and it is not always clear why the CPU is blocked and hence the most appropriate mitigation. Bearing in mind OoO processors, in conjunction with modern memory controllers, can issue a non-blocking retrieve from main memory and, whilst this is on-going, keep themselves busy executing later, non-dependent instructions, out of order, cache misses by themselves are not necessarily problematic. 

Instead one is concerned with \emph{delinquent loads}, where the CPU is forced to stall due to the cache miss and, from a micro-architecture perspective, there are three reasons why the CPU could stall when a cache miss occurs \cite{day-life-miss}. Firstly, structural blockages are when the CPU micro-architecture simply runs out of resources, such as slots in the reorder buffer or physical registers. The second reason for delinquent loads are data blockages which is where instructions that depend on the loaded data, either directly or indirectly, begin to pile up in the reservation station and eventually the reservation station is full and the CPU stalls, effectively because all instructions which it could issue and execute out of order have some form of dependency on the data that is being retrieved from main memory. The third reason that a cache miss can cause the CPU to stall is due to control dependencies. This is where branches are dependant on the loaded data and the CPU has miss predicted a branch due to the missing data. Once an incorrect branch prediction is recognised, all instructions after the branch will be flushed once the delinquent load completes and work done by the CPU whilst the load was in-progress wasted.

It is important to understand the underlying causes and nature of these delinquent loads, because approaches such as software prefetching, which fetches data from main memory into cache before it is needed, adds additional instructions and takes up significant amounts of memory bandwidth. As such, with modern OoO processors, the indiscriminate use of software prefetching in user code can actually reduce performance \cite{prefetch-work-not}. Based on detailed profiling of the code it was this single procedure that is by far most impacted by the irregularity of memory access. From Figure \ref{fig-poor-counters-irregular} it can be seen that stalls due to resource limits in the store buffer were very significant. Once a write is completed it is actually written to the micro-architecture's store buffer and the CPU will continue. From the store buffer the value is written to the corresponding location in L1 cache, with fetching from a lower level cache or main memory in to L1 cache performed if needed. On the Ivy Bridge micro-architecture there are only 36 possible entries in the store buffer (in contrast to 64 in the load buffer) and in this situation locations that are not in the L1 cache are being written to so fast that the store buffer is becoming full and the CPU stalling due to this structural blockage. From experience, you have to be a little careful here as quite often the store buffer, with its smaller number of entries than the load buffer, fills up first and so shows a high percentage of the overall stalls. If this alone is fixed, then then often the loading of data then shows a similarly high number of stalls, as the underlying problem applied to both the storing and loading of values but effectively the stalling of the loads was hidden by the stalling on the stores.

To this end we adopted the technique of software pipelining \cite{sw-pipeline}, in conjunction with software prefetching for both the write and read of Listing \ref{lst:irregularaccess}. This is where the code runs in a pipelined fashion and required memory location is prefetched, non-blocking, ahead of time. The idea is that when it comes to using the data, that is already in the cache and no external memory access need be issued. This is illustrated in Listing \ref{lst:swpipelined} and, once inside the outer loop, we start off by prefetching the first \emph{PREFETCH\_DISTANCE} elements in the \emph{matrix} and \emph{equations} arrays for the value of \emph{j} (lines 2 to 5). Once this has completed, we then go into the inner loop, and each iteration of this inner loop first issues non-blocking prefetch calls for the \emph{matrix} and \emph{equations} values \emph{PREFETCH\_DISTANCE} ahead of \emph{i} (lines 8 to 11). After this, in that same iteration of the inner loop, we then write to and read from those already prefetched variables at the \emph{i}'th index (line 12). The idea being that, as we are working with these arrays at line 12, the elements at \emph{i} have already been fetched and will be served from the cache. Prefetching doesn't impact the correctness at all, so the initial prefetching for the outer loop at lines 2 to 5 is optional, and we could do without it, but this does improve performance because the initial values of \emph{i} are being prefetched by the memory controller whilst further prefetching calls are being issued for larger values of \emph{i} up to \emph{PREFETCH\_DISTANCE}.

\begin{lstlisting}[frame=lines,caption={Illustration of software pipelined procedure for pre-fetching},label={lst:swpipelined},numbers=left]
do j=1, n
    do i=1, PREFETCH_DISTANCE
        call do_prefetch(matrix(dataloc(i), j))
        call do_prefetch(equations(inputdata(i,j)))
    end do
    do i=1, n
        k=i+PREFETCH_DISTANCE
        if (k .le. n) then
            call do_prefetch(matrix(dataloc(k), j))
            call do_prefetch(equations(inputdata(k,j)))
        end if
        matrix(dataloc(i), j)=equations(inputdata(i,j)) + ....
    end do
end do
\end{lstlisting}

This is called software pipelining because effectively we have a two stage pipeline, the first stage running \emph{PREFETCH\_DISTANCE} ahead of \emph{i} and prefetching data, and the second stage running at \emph{i} and taking advantage of the prefetched data. In fact, the value of \emph{PREFETCH\_DISTANCE}, effectively the gap between the prefetch and the memory access, is very important \cite{prefetch-work-not}. If this is too small then prefetching is ineffective because the memory access has not yet completed and the main memory access at line 12 would still block, if it is too large then there is a danger that the prefetched data will be flushed from cache before it is used. From profiling, we found that the optimal \emph{PREFETCH\_DISTANCE} was 16 on ARCHER, although this will vary from architecture to architecture. 

The \emph{do\_prefetch} subroutines that we call in Listing \ref{lst:swpipelined} uses the ISO C bindings to wrap Intel's \emph{\_mm\_prefetch} function and we instruct this call to prefetch into the L1 cache. The results of profiling the software pipelining approach is illustrated in Figure \ref{fig-fixed-counters-irregular}, where we can see that the total number of instructions has increased dramatically for this procedure, over three times, but crucially the total number of cycles has more than halved. In contrast to the non-prefetching approach which averaged 2.26 cycles per instruction, here our CPI is 0.39 (lower is better), and on average 2.53 instructions are executed every clock cycle (higher is better). This is much closer to Ivy Bridge's theoretical maximum of 5 instructions per clock cycle. From comparing Figures \ref{fig-poor-counters-irregular} and \ref{fig-fixed-counters-irregular}, it can be seen that the number of cycles stalled is far less and the number of L1 cache hits is four times greater. As such this has reduced the percentage run time of this procedure by almost three times.

At only two stages, our software pipeline in Listing \ref{lst:swpipelined} is fairly simple. It would of-course be possible to extend this to a third stage and prefetch the \emph{inputdata} and \emph{dataloc} arrays also. However, from following a similar profiling investigation we found that, because accesses into these structures follow a contiguous approach, they already make good use of the cache and the added instruction count results in a decrease of overall performance.

\begin{figure}
\begin{center}
\begin{tabular}{ | c | c | }
\hline
Counter description & Value \\ \hline
Number of cycles & 32,834,830,469 \\
Number of instructions & 83,862,040,343\\
Total resource stalls & 1,903,905,993\\
Resource stalls (store buffer) & 629,838,931\\
Cycles no instructions are issued & 1,927,696,028\\
L1 cache hits & 19,907,028,904\\
\hline
Instructions per cycle  & 2.53\\
Cycles per instruction & 0.39 \\
Total \% cycles stalled & 5.8 \\
\hline
\end{tabular}
\end{center}
\caption{Hardware counter values using prefetching via software pipelining approach for mitigating irregular memory access}
\label{fig-fixed-counters-irregular}
\end{figure}


\section{Solving the normal equations using PETSc and SLEPc}
\label{sec:slepcwork}
Once the normal equations have been built, the next stage is to then solve them. The code finds all the eigenpairs (eigenvalues and associated eigenvectors) of the matrix and then, for eigenvalues larger than a threshold, applies the eigenpair to the RHS in combination with some weight. Over 99\% of the run time of the previous model is in calculating the eigenpairs of the matrix, with the application of these once found to the RHS being trivial. This requirement of finding all the eigenpairs is somewhat unusual, with many models just focusing in one area of the spectrum. Actually calling into SLEPc and PETSc, from a code perspective, was fairly trivial to code up once the normal equations data structures were distributed correctly. When the code begins we initialise these frameworks using the appropriate API calls and construct the distributed data-structures, it is the PETSc library that determines the decomposition and provides the start and end matrix rows for each process. 

\begin{lstlisting}[frame=lines,caption={Sketch of PETSc/SLEPc integration},label={lst:petscslepc},numbers=left]
call MatCreate(solving_communicator, A, ierr)
call MatSetType(A, MATMPIAIJ, ierr)
call MatSetSizes(A, PETSC_DECIDE, PETSC_DECIDE, num_bases_petsc_int, num_bases_petsc_int, ierr)
call MatSetFromOptions(A, ierr)
call MatSetUp(A, ierr)

call MatGetOwnershipRange(A,matrix_start,matrix_end,ierr)
.....
call MatMPIAIJSetPreallocation(A, matrix_num_rows, PETSC_NULL_INTEGER, num_of_bases-(matrix_end-matrix_.start), PETSC_NULL_INTEGER, ierr)

call MatSetValues(A, matrix_num_rows, start_row_index, num_columns, start_col_index, matrix_values, INSERT_VALUES, ierr)

.....
call EPSCreate(solving_communicator, eps, ierr)
call EPSSetOperators(eps, A, PETSC_NULL_MAT, ierr)
call EPSSetDimensions(eps, num_bases_petsc_int, PETSC_DEFAULT_INTEGER, PETSC_DEFAULT_INTEGER,  ierr)
call EPSSetFromOptions(eps, ierr)

call EPSSolve(eps, ierr)
call EPSGetConverged(eps, num_found, ierr)
do i=1,num_found
	call EPSGetEigenvalue(eps, i-1, eigr, eigi, ierr)
	.....
end do
\end{lstlisting}

Listing \ref{lst:petscslepc} illustrates a sketch of the code used to integrate PETSc and SLEPc with the model. Between lines 1 and 5 the matrix is created to be of type \emph{MPIAIJ}, which tells PETSc that this is a parallel matrix decomposed over the MPI communicator specified at line 1. At line 7 we determine the global index of the starting and ending row held on the local process, and at line 8 memory required for each processes' portion of the matrix data structure is allocated. This preallocation is an optimisation option, and we found that it was very important for performance when setting values of the matrix at line 10, as otherwise PETSc allocated memory lazily which incurs significant overhead. Until this point in the code, the calls are to the PETSc library directly, and at line 14 we call into the SLEPc library for the first time to create an Eigen Problem Solver (EPS) for matrix \emph{A} (line 15). The \emph{EPSSetDimensions} call at line 16 determines how many eigenvalues to find, and the eigen-solve itself is called at line 19. The number of eigenvalues that the solver actually found is deduced at line 20, and then between lines 21 and 24 we extract each eigenvalue and perform work on it which also involves extracting each processes's portion of the eigenvector using the {EPSGetEigenvector} call which is omitted for brevity.

As described in Section \ref{sec:petscslepc}, SLEPc provides numerous Eigen solvers but after extensive testing we found that, for this problem the default, Krylov-Schur, solver is most accurate, with respect to the previous code whose highly stable Givens reduction was taken as a base line, and fastest too. This illustrates one of the benefits of the SLEPc and PETSc libraries, as from a code perspective it is trivial to experiment with different solvers and configurations. The most important configuration option we found was to inform SLEPc that the problem is generalized Hermitian. Effectively it makes the problem easier to solve and is important because, not only does this significantly improve performance in comparison with the default non-Hermitian approach, but also the previous model, which we take as the ground truth, makes this assumption and-so the models most closely match in this configuration. We also experimented with the maximum projected dimension, which trades off memory usage for redundant computation when running in parallel. However, this did not make a significant difference to performance or memory usage.

This raises an important point however, as the MEME model itself is generic in terms of the specifics of the problem. Future users, interested in different problems, will inject their own procedures in for generating the values that feed into the building of the normal equations. Therefore it is likely that future users of the code, with their different equations could benefit from SLEPc solvers and options different to the ones optimal to this problem. It is important that, the first instance, these individuals can experiment with different PETSc and/or SLEPc options via command line arguments. 

\section{Performance and scaling}
\label{sec:performance}
\begin{figure}
	\begin{center}
		\includegraphics[scale=0.30]{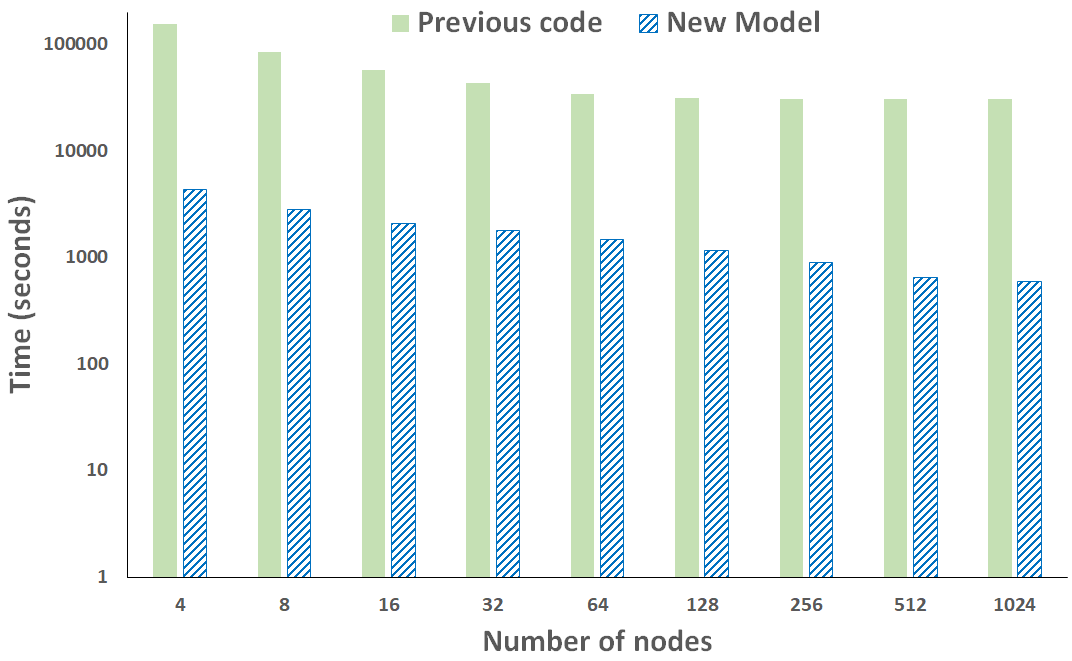}
	\end{center}
	\caption{Performance comparison of previous and new model with system size of 10,000 coefficents and 4.3 million data points. The new model is 51 times faster at 1024 nodes than the previous model.}
	\label{fig-perf-scaling}
\end{figure}

Performance and scaling experiments have been carried out on ARCHER, a Cray XC30. Nodes have two twelve core Ivy Bridge processors and most commonly have 64GB RAM. All experiments have been compiled with the Cray compiler version 8.6.5, PETSc version 3.8.4 and SLEPc version 3.8.3. In this section we concentrate on a system size of 10,000 coefficients and 4.3 million input data items, contrasting the performance of our new code using PETSc and SLEPc for solving the normal equations, against the previous model. The results of this experiment are illustrated in Figure \ref{fig-perf-scaling} (run time, on the vertical, is log scale), where it can be seen that the performance of the newly developed model is very significantly faster than that of the previous code up to 1024 nodes (24576 cores) and scales far better.

\begin{figure}
	\begin{center}
		\includegraphics[scale=0.30]{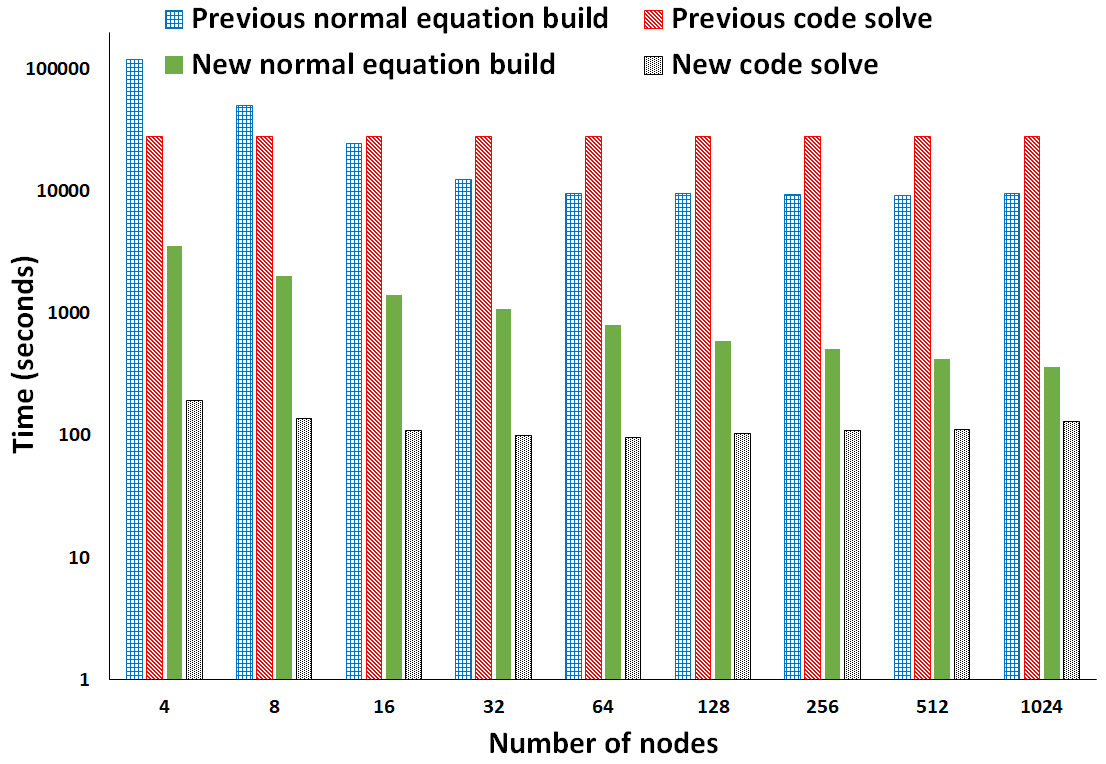}
	\end{center}
	\caption{Amount of run time for solver and building of normal equations for previous and new code as the number of nodes is scaled. Strong scaling effects also play a part here, but at the optimal number of nodes (64) the SLEPc approach is 294 times faster than the previous model's bespoke solver and at 1024 nodes our new normal equation building is 26 times faster than the previous model's approach. }
	\label{fig-reason-performance}
\end{figure}

Figure \ref{fig-reason-performance} illustrates the breakdown of timings for the results in Figure \ref{fig-perf-scaling} between the normal equation build time and the solver time (again, vertical axis run time is log scale). It can be seen that, as expected, the solver time of the previous code is constant at 28,500 seconds irrespective of the parallelism due to its sequential nature. The run times for the normal equation building in the new code, along with the solve are significantly smaller than that of the previous code. From Figure \ref{fig-reason-performance} it can be seen that the most significant difference in terms of performance between the two models is in the solver time, where our new code takes 194 seconds over 4 nodes and 97 seconds over 64 nodes in contrast to 28,500 seconds for the previous model regardless of parallelism. This represents a speed up for 294 times and illustrates one of the major benefits to using SLEPc and PETSc over the bespoke serial solver in the previous code, both in terms of raw computational performance and also the ability to leverage parallelism. In terms of solver performance, 97 seconds over 64 nodes was the fastest that it ran, with the solver run time slightly increasing beyond this point, e.g. to 130 seconds over 1024 nodes. This is because, with 10,000 model coefficients, we are hitting the limits of strong scaling and the problem size was not big enough to take advantage of the increased parallelism.

The building of the normal equations is also substantially faster in the new code compared to the previous code, 26 times at 1024 nodes. This is a combination of the parallel and serial optimisations as described in Section \ref{sec:matrixandrhs}. Whilst the run time of this aspect of the code does continue to drop as the number of nodes increases, with 365 seconds over 1024 nodes being the fastest, again we are hitting the limits of strong scaling here as there are diminishing returns as one gets beyond 128 nodes (3072 cores on ARCHER).

\subsection{Result accuracy}
\label{sec:accuracy}
Figure \ref{fig-results-difference} illustrates the percentage difference between results generated by the previous code and the new model in the experiment of this section. Because we have changed so significantly how the normal equations are built and solved, there was a significant question around how the models would compare. Going into this project we had a target of 0.1\% mean difference between the results of the previous and new model. 

Figure \ref{fig-results-difference} contains the minimum, maximum and mean difference for each different result file generated by the code (Ffit to XYZfit\_c(3)) and it can be seen that the mean difference is well below this 0.1\% difference target. The Ffit and Fobsfit results contain one result for each element, whereas for the XYZfit, XYZobsfit and XYZfit\_c results there are three separate result elements. In the later case we only report elements 1 and 3 in Figure \ref{fig-results-difference} because element 2 is exactly equal, a zero percent difference, throughout all the results between the two models.

\begin{figure}
\begin{center}
\begin{tabular}{ | c | c c c| }
\hline
Metric & \makecell{Minimum \\ \%difference} & \makecell{Maximum \\ \%difference} & \makecell{Mean \\ \%difference} \\ \hline
Ffit & 0.000019 & 0.000053 & 0.000026\\  
Fobsfit & 0.000014 & 0.000029 & 0.000019 \\
XYZfit(1) & 0.000010 & 0.034425 & 0.000091 \\
XYZfit(3) &0.000010 & 0.312869 & 0.000204 \\
XYZobsfit(1) & 0.000010 & 0.000081	& 0.000033 \\
XYZobsfit(3) & 0.000010 & 0.182673	& 0.000130 \\
XYZfit\_c(1) & 0.000010	& 0.007966	& 0.000073 \\
XYZfit\_c(3) & 0.000010	& 0.011187	& 0.000085 \\
\hline
Model coefficients & 0 & 43 & 0.048114\\
\hline
\end{tabular}
\end{center}
\caption{Percentage result difference between results generated by previous model and new model with experiment of figure 12 over 32 nodes. Whereas the maximum difference of coefficients seems large, these were all very tiny numbers and as such had no signifiant impact on the overall results.}
\label{fig-results-difference}
\end{figure}

The model coefficients of Figure \ref{fig-results-difference} are slightly different to all the other result file entries in the table. These are the raw generated coefficients that come directly from the eigenvalues and vectors, before they are then applied to the input data to generate the results. The geomagneticists use them as a sanity check to ensure the results are consistent and whilst a maximum difference of 43\% seems very significant (especially in comparison to the 0\% minimum and 0.048\% mean), it should be noted that all the differences in this large range represent very tiny numbers smaller than 1e-30. After discussion with the geomagneticists it was determined that, whilst percentage wise these differences might seem significant, in reality because those numbers are so tiny they make no significant difference to the overall results.

Generally speaking it surprised us how closely the results match between the previous model's Givens reduction approach and SLEPc. One of the strong reasons for the Given's reduction was its stability and perceived accuracy, and we have shown that actually this can be replaced by a much faster, parallel method, which has no qualitative impact on the accuracy of the overall results. This is an important point because there are significant advantages to SLEPc and, in this context, we can benefit from these without sacrificing any degree of accuracy which the geomagneticists were worried about.

\section{Scaling the number of model coefficients}
\label{sec:largecoefs}
A major limit of the current code was it's inability to scale beyond 10,000 model coefficients due to memory limitations. Having decomposed the matrix and RHS data structures across processes this limitation has been mitigated. However the direct eigen-solve in SLEPc requires the allocation of a significant data structures during this direct solve on each process, especially when searching for all eigenvalues and vectors as we require here. Whilst there are options for experimenting with the maximum projected dimension (which trades off memory usage and computational recalculation), and replacing the solver with the MUMPS parallel direct linear solver, from experimentation we found that these had no impact on the very large memory usage of SLEPc.


\begin{figure}
\begin{center}
\begin{tabular}{ | c | c | }
\hline
\makecell{Number of \\ coefficients} & \makecell{Direct solve size (MB) \\ per process} \\ \hline
10000 & 1498 \\
20000 & 9543 \\
30000 & 23908 \\
40000 & 38200 \\
50000 & 57300 \\
60000 & 85950 \\
70000 & 114600 \\
\hline
\end{tabular}
\end{center}
\caption{Memory usage requirements, per process, of direct solver by the number of coefficients}
\label{fig-mem-usage-needs}
\end{figure}

Figure \ref{fig-mem-usage-needs} illustrates how the memory usage of the direct solver grows as the number of model coefficients is increased. This memory requirement is in addition to other data structures including the matrix and RHS, and is not parallelisable. Therefore  irrespective of the number of processes, this amount of memory must be allocated per process and is a very serious limitation. Bearing in mind most nodes in ARCHER, the Cray XC30 used for this work, have 64GB RAM in total and 32GB per NUMA region, ways round this needed to be found. Whilst there is a trend to increase the amount of memory per node with more modern machines, memory usage is still a limitation however as we reach large coefficient sizes.

\subsection{Hybridising the code with OpenMP to address the memory challenge}
Based on the experiments done in Section \ref{sec:performance}, the building of the normal equations well as we increase the amount of parallelism. However, the direct solve does not scale quite so well and experiments on over 16 nodes in Figure \ref{fig-perf-scaling} only decreases the run time of the solver slightly, or even increases it. As the memory limits of the direct solver are on a process by process basis, we decided to hybridise the code using OpenMP to reduce the number of processes running per node. Whilst there has been some work done in PETSc to support hybrid OpenMP/MPI parallelism \cite{hybrid-petsc} this is not particularly mature and as yet is not included in the main PETSc distribution. Therefore we just run the solve on a process by process basis. With OpenMP applied throughout our own code, aspects such as the building of the normal equations takes advantage of all the cores using thread level concurrency, but the solving of these does not. As described in Section \ref{sec:petscslepc}, an advantage of PETSc is that it is trivial to modify configuration options and as such in the future it will be trivial to change the form of parallelism by simply selecting a different type of matrix data structure, if and when hybrid support becomes more mature.

\begin{lstlisting}[frame=lines,caption={Illustration hybridised OpenMP/MPI kernel},label={lst:hybridised},numbers=left]
!$omp parallel do private(j, dw, il) reduction(+:rhs_norm_equations)
    do j=1, lines_in_y
      ....
      do il=1, number_of_rows
        rhs_norm_equations(il)=rhs_norm_equations(il)+.....
      end do
    end do
!$omp end parallel do
\end{lstlisting}

Listing \ref{lst:hybridised} sketches the hybrid OpenMP code for one of the kernels involved in building the normal equations. All OpenMP calls are loop based directives, and in this instance we are performing an OpenMP reduction between the threads once local iterations of the loop have completed. As before, the matrix and RHS data structures are decomposed on a process by process basis and when building the normal equations, the input data is split up between the OpenMP threads. Each thread works on the processes' data structure chunk but with a separate subset of the input data to process. This is important, because it means we can start MPI in \emph{funneled} threading mode rather than the much slower \emph{multiple} mode. It should also be noted that, due to the non-associativity of floating point arithmetic, using OpenMP reductions can generate slightly different results from one threading level to another. However, this is still well within tolerance and we also provide additional code to perform the reduction in guaranteed order, via local variables and explicit addition but with some minor degradation in performance, if this is critical to the end user. 


\begin{figure}
\begin{center}
\begin{tabular}{ | c | c c c c | }
\hline
Nodes & \makecell{2 threads \\ per process} & \makecell{4 threads \\ per process} & \makecell{6 threads \\ per process} & \makecell{12 threads \\ per process} \\ \hline
40 & 119.02 & 88.05 & 76.94 & 66.76 \\
36 & 118.62 & 88.89 & 76.42 & 71.27 \\
27 & 117.41 & 90.94 & 81.96 & 77.73 \\ 
18 & 124.13 & 85.55 & 92.63 & 83.97 \\
9 & 140.69 & 120.6 & 120.51 & 138.5 \\ 
5 & 208.76 & 190.75 & 199.14 & 244.41 \\ 
3 & 369.42 & 317.71 & 352.72 & 414.8 \\ 
\hline
\end{tabular}
\end{center}
\caption{Total run time changes with respect to threads per process based on 500,000 data points over 8000 coefficients}
\label{fig-total-openmp}
\end{figure}

\begin{figure}
\begin{center}
\begin{tabular}{ | c | c c c c | }
\hline
Nodes & \makecell{2 threads \\ per process} & \makecell{4 threads \\ per process} & \makecell{6 threads \\ per process} & \makecell{12 threads \\ per process} \\ \hline
40 & 42.99 & 29.52 & 24.59 & 20.33 \\
36 & 44.19 & 30.69 & 25.86 & 21.72 \\
27 & 48.61 & 34.39 & 29.69 & 25.96 \\
18 & 56.05 & 41.82 & 37.28 & 34.49 \\
9 & 78.95 & 64.17 & 60.55 & 59.72 \\
5 & 127.4 & 109.47 & 104.34	& 111.05 \\
3 & 245.82 & 189.65 & 195.11 & 190.12 \\
\hline
\end{tabular}
\end{center}
\caption{Matrix and RHS building time changes with respect to threads per process based on 500,000 data points over 8000 coefficients}
\label{fig-matrix-rhs-building-openmp}
\end{figure}

\begin{figure}
\begin{center}
\begin{tabular}{ | c | c c c c | }
\hline
Nodes & \makecell{2 threads \\ per process} & \makecell{4 threads \\ per process} & \makecell{6 threads \\ per process} & \makecell{12 threads \\ per process} \\ \hline
40 & 55.7 & 44.8 & 40.29 & 36.68 \\ 
36 & 54.67 & 43.76 & 38.72 & 39.71 \\ 
27 & 49.32 & 42.07 & 40.4 & 41.63 \\
18 & 48.04 & 29.84 & 43 & 39.62 \\
9 & 42.07 & 42.19 & 48.01 & 68.15 \\
5 & 61.72 & 66.64 & 82.3 & 122.34 \\
3 & 103.36 & 113 & 144.35 & 212.99 \\
\hline
\end{tabular}
\end{center}
\caption{Solver time changes with respect to threads per process based on 500,000 data points over 8000 coefficients}
\label{fig-solver-openmp}
\end{figure}

We found that utilising threading made a difference to overall performance with the model and ran a smaller experiment than those detailed in Section \ref{sec:performance}, with 500,000 input data items and 8000 model coefficients. Figure \ref{fig-total-openmp} illustrates how the run time of the model changes with respect to the number of threads per process for this experiment. For each row, for instance the 40 nodes row, the overall number of cores remains unchanged (960 in this case), and with two threads per process we have 480 process each running over 2 threads, out to twelve threads per process resulting in 80 processes each with 12 threads. It isn't as simple as saying one level of threading is \emph{better} or \emph{worse} than another. Instead, it depends heavily on the overall amount of parallelism (number of nodes) and it can be seen that when using 40 nodes the run time can be almost halved by the simple configuration change of going from 2 threads per node to 12 threads per node. However over three nodes going from 2 to 12 threads per node reduces performance by almost a third.

To break it down further, Figures \ref{fig-matrix-rhs-building-openmp} and \ref{fig-solver-openmp} show the run time, for each configuration, for building the normal equations, and the solver time respectively. Based on the solver run times it can be seen that we are hitting the limits of strong scaling with this experiment size as, at small node sizes it is advantageous to use more parallelism and at larger numbers of nodes less, for this problem size this factor dominates at smaller numbers of nodes. For the normal equation building, irrespective it is advantageous to use more threading where possible and this is because it reduces the amount of MPI communications needed, with this problem size this factor dominates at larger numbers of nodes. As all processes potentially need to communicate with all other processes, this can be an expensive. 
\begin{figure}
	\begin{center}
		\includegraphics[scale=0.45]{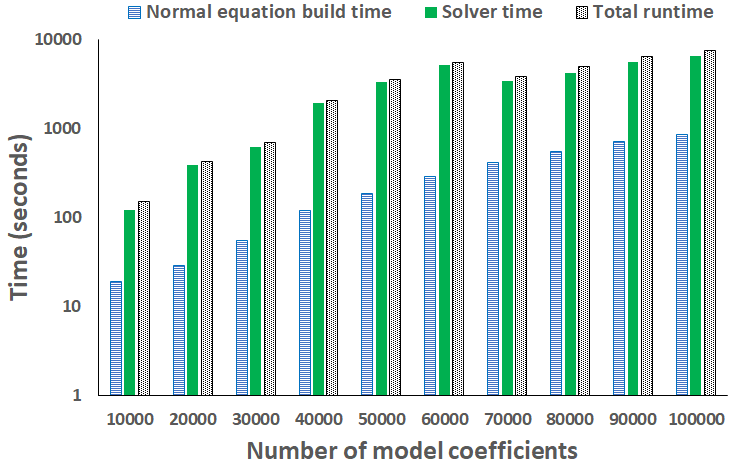}
	\end{center}
	\caption{Scaling the number of model coefficients over 40 nodes of ARCHER}
	\label{fig-large-scaling}
\end{figure}

Based on this hybridisation we can now run the new model with problem sizes much larger than the 10,000 coefficients that the previous model was limited to. Figure \ref{fig-large-scaling} illustrates an experiment, over 40 nodes of ARCHER, as we scale the number of model coefficients up to 100,000. There are a number of caveats and noteworthy points to be highlighted here. It can be seen that there is a sharp increase in run time at 40,000 coefficients and this is where we had to switch, due to memory limits, from a process per NUMA region to a process per node in order to fit into memory. At 50,000 coefficients we started using the large memory nodes on ARCHER, which contain 128GB of RAM, and provided some extra headroom for these runs. In terms of modern computing, 128GB of RAM isn't a particularly large amount, and once we reached 70,000 coefficients this memory was exhausted. In order to model problem sizes of 70,000 coefficients and above, we split the problem in two when it came to eigen-solving, first finding the \emph{n/2} largest eigenvalues in magnitude and applying these and their corresponding eigenvectors to the RHS, and then finding the \emph{n/2} smallest eigenvalues in magnitude and applying these and their corresponding eigenvectors to the RHS. This worked well, and actually resulted in a slight decrease in run time because at that stage two smaller solves was faster than one larger one. 

It should be noted that we tried a number of approaches to finding different portions of the spectrum, such as those closest to a target and then iteratively increasing the target. None of these worked very well and the only reliable approach was to split the solve in two and finding the \emph{n/2} largest and \emph{n/2} smallest eigenpairs. In our mind this is the current limitation of our model and, as it currently stands, when the problem size reaches a point where the two separate solves run out of memory, most likely at around 120,000 coefficients on ARCHER further investigations will need to be performed. There are a number of points to bear in mind here, in addition to the obvious point that larger amounts of memory are becoming much more commonplace which mitigate this issue somewhat. Firstly we are finding all the eigenvalues and then throwing away the smallest ones beneath a threshold. At 100,000 model coefficients this represents a large number that are being discarded and, as the direct solver memory usage is determined by the number of eigenvalues being searched, we could likely reduce the number of eigenvalues being searched because of the threshold and improve the situation. Secondly, as we reached 70,000 coefficients PETSc had to be compiled with 64 bit matrix indices, this isn't a problem in itself but does illustrate the significant size of the matrix and general problem we are working with here.

\subsection{An iterative vs direct solver approach}
We have described the limitations of using SLEPc to find eigenpairs for large numbers of model coefficients. An alternative approach is to move from an eigen-based direct solve to an iterative solver and as our solver is already calling into PETSc, this is fairly easy to do from a code perspective. In fact, in terms of the code, the only change required was to initialise a PETSc iterative solver via the \emph{KSP} data structure and hook this up to our matrix and RHS. 

\begin{figure}
		\includegraphics[scale=0.40]{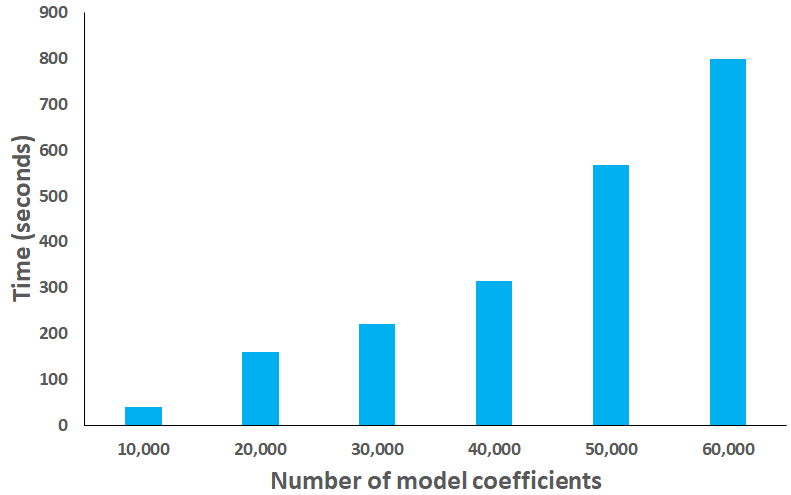}
	\caption{Performance of the iterative solver as the number of model coefficients is scaled, running on 40 nodes of ARCHER}
	\label{fig-iterative-solver}
\end{figure}

Using the ILU preconditioner and GMRES solver, with a relative tolerance of 1e-4, Figure \ref{fig-iterative-solver} illustrates the run time of the iterative solver over 40 nodes of ARCHER as we scale the problem size. As the iterative solver approach does not suffer from the memory limitations of the direct solver approach, we are able to run one process per core (e.g. 960 processes) for this experiment and that is the configuration which was used. It can be seen that the run time here is considerably less than that of the direct solver, for instance 800 seconds iterative solver time verses 5000 seconds direct solver time in Figure \ref{fig-large-scaling} for a problem size of 60,000 coefficients. We also have many more options, for instance being able to experiment with the accuracy and it is highly likely that different users of the MEME model, with their different problems, would require different levels of accuracy.

One of the disadvantages of the iterative approach however is its stability. Up to 10,000 coefficients the results match fairly closely between an iterative and direct solver approach, within around 5\%. Whilst this is outside the accuracy limit that the geomagneticists will currently accept, they believe that with some further work this could be fairly easily reduced and confidence provided in the results. However, we found that as we scale the number of model coefficients beyond this point, the solution becomes less and less accurate and after 60,000 model coefficients the solver starts to diverge. Clearly there is further work required here to fully understand the impact that the iterative solver is having on the system and optimal configuration settings.

\section{Conclusion and further work}
\label{sec:conclusions}
In this paper we have explored the use of PETSc and SLEPc in solving a system of normal equations for modelling the earth's magnetic field. We have demonstrated significant benefits to replacing the bespoke eigen-solver of the previous MEME model with that provided by SLEPc, and have shown that, at no qualitative different to the output result accuracy, very significant improvements in performance are possible by leveraging this popular library. In terms of problem size, the new model is capable of modeling systems very significantly larger than those that were previously attainable. However when it comes to large numbers of coefficients, and more generally using SLEPc for finding very many eigenpairs in a large system, there are caveats and limitations around memory usage and a number of mitigations are required to work around these. ARCHER, the XC30 used for this work is over five years old now and inevitably more modern machines with larger amounts of memory will be impacted less by this, but still it is important to bear in mind as scientific ambitions of this model and use of SLEPc in general will only continue to grow.

Adopting PETSc and SLEPc has had implications far beyond the specific solver itself and we have also described our approach for building the symmetric matrix of normal equations in a distributed fashion. We demonstrated an approach that requires minimal process coordination, gives reasonable load balance and avoids any replication of computation. Whilst inter-node performance was a major aspect of this, we also found that single core performance was significantly limited by the irregularity of memory access. Careful profiling helped us understand the problem further and this was then addressed by the use of software prefetching in conjunction with software pipelining. We demonstrated significant benefits to both these approaches and combined the building of the normal equations is now very significantly faster than in the previous model. 

In this paper we have briefly touched on the topic of solving the normal equations via an iterative approach rather than a direct eigen-solver. Bearing in mind the challenges around SLEPC's memory requirements, this is an important avenue of further investigation. It is our feeling that, whilst we did modify the normal equations slightly to make them applicable to an iterative solver, more generally this is still unstable and a more fundamental rethink from the geomagneticists is required to take full advantage of an iterative solver approach. Nevertheless, we have demonstrated that there is potential benefit to doing this and, from a code perspective, it is trivial to switching out the direct solver and replacing it with an iterative solver.

The use of software prefetching in HPC is still an immature topic, but it is a useful way of addressing the irregularity of memory accesses and has the potential to be an important future approach. A major limitation to this, as it currently stands, in our new MEME model is the use of the \emph{PREFETCH\_DISTANCE} internal to the code. This is because the optimal value depends upon the micro-architecture and as such, profiling is needed on each new machine to provide performance portability of the model. Previous work \cite{prefetch-distance-dynamic} has been done on dynamically tuning this prefetching distance in an HPC code, and we believe that this would be an interesting and important avenue of further research in the MEME model. It is also likely that, as the code executes, the optimal prefetching distance will changed dynamically and as-such this approach could result in improved performance more generally.

\section*{Acknowledgments}
This work was funded under the embedded CSE programme of the ARCHER UK National Supercomputing Service (http://www.archer.ac.uk)


\begin{thebibliography}{1}

\bibitem{igrf}
Thébault, E., Finlay, C.C., Beggan, C.D., Alken, P., Aubert, J., Barrois, O., Bertrand, F., Bondar, T., Boness, A., Brocco, L. and Canet, E., 2015. International geomagnetic reference field: the 12th generation. Earth, Planets and Space, 67(1), p.79.

\bibitem{wmm}
Chulliat, A., Macmillan, S., Alken, P., Beggan, C., Nair, M., Hamilton, B., Woods, A., Ridley, V., Maus, S. and Thomson, A., 2015. The US/UK world magnetic model for 2015-2020.

\bibitem{meme}
Brown, W., Beggan, C. and Macmillan, S., 2016. Geomagnetic jerks in the SWARM era.

\bibitem{slepc-cfd}
Garnaud, X., Lesshafft, L., Schmid, P.J. and Chomaz, J.M., 2012. A relaxation method for large eigenvalue problems, with an application to flow stability analysis. Journal of Computational Physics, 231(10), pp.3912-3927.

\bibitem{slepc-material-science}
Dettori, R. and Colombo, L., 2018. THERMAL PROPERTIES OF TPD-BASED ORGANIC GLASSES. Istituto Lombardo-Accademia di Scienze e Lettere-Incontri di Studio.

\bibitem{slepc-structural}
Genoese, A., Genoese, A., Bilotta, A. and Garcea, G., 2014. Buckling analysis through a generalized beam model including section distortions. Thin-Walled Structures, 85, pp.125-141.

\bibitem{slepc-earth}
Marti, P., Calkins, M.A. and Julien, K., 2016. A computationally efficient spectral method for modeling core dynamics. Geochemistry, Geophysics, Geosystems, 17(8), pp.3031-3053.

\bibitem{slepc-geomag}
Vidal, J. and Schaeffer, N., 2015. Quasi-geostrophic modes in the Earth's fluid core with an outer stably stratified layer. Geophysical Journal International, 202(3), pp.2182-2193.

\bibitem{elpa}
Marek, A., Blum, V., Johanni, R., Havu, V., Lang, B., Auckenthaler, T., Heinecke, A., Bungartz, H.J. and Lederer, H., 2014. The ELPA library: scalable parallel eigenvalue solutions for electronic structure theory and computational science. Journal of Physics: Condensed Matter, 26(21), p.213201.

\bibitem{petsc}
Balay, S., Gropp, W.D., McInnes, L.C. and Smith, B.F., 1997. Efficient management of parallelism in object-oriented numerical software libraries. In Modern software tools for scientific computing (pp. 163-202). Birkhäuser, Boston, MA.

\bibitem{slepc}
Hernandez, V., Roman, J.E. and Vidal, V., 2005. SLEPc: A scalable and flexible toolkit for the solution of eigenvalue problems. ACM Transactions on Mathematical Software (TOMS), 31(3), pp.351-362.

\bibitem{eigen}
Guennebaud, G., Jacob, B., 2010. Eigen v3. [Online]. [10 April 2019]. Available from: http://eigen.tuxfamily.org

\bibitem{eigen-usage}
Springer, P., Su, T. and Bientinesi, P., 2017, June. HPTT: a high-performance tensor transposition C++ library. In Proceedings of the 4th ACM SIGPLAN International Workshop on Libraries, Languages, and Compilers for Array Programming (pp. 56-62). ACM.

\bibitem{mpi-message-size}
Balaji, P., Chan, A., Gropp, W., Thakur, R. and Lusk, E., 2008, September. Non-data-communication overheads in MPI: analysis on Blue Gene/P. In European Parallel Virtual Machine/Message Passing Interface Users’ Group Meeting (pp. 13-22). Springer, Berlin, Heidelberg.

\bibitem{day-life-miss}
Karkhanis, T. and Smith, J.E., 2002, May. A day in the life of a data cache miss. In Workshop on Memory Performance Issues (Vol. 99).

\bibitem{prefetch-work-not}
Lee, J., Kim, H. and Vuduc, R., 2012. When prefetching works, when it doesn’t, and why. ACM Transactions on Architecture and Code Optimization (TACO), 9(1), p.2.

\bibitem{sw-pipeline}
Sánchez, J. and González, A., 1999. Software data prefetching for software pipelined loops. Journal of Parallel and Distributed Computing, 58(2), pp.236-259.

\bibitem{prefetch-distance-dynamic}
Hadade, I., Jones, T.M., Wang, F. and di Mare, L., 2018, November. Software prefetching for unstructured mesh applications. In 2018 IEEE/ACM 8th Workshop on Irregular Applications: Architectures and Algorithms (IA3) (pp. 11-19). IEEE.

\bibitem{givens}
Chen, S., 2014. Reduction of a Symmetrical Matrix to Tridiagonal Form on GPUs.

\bibitem{givens-slow}
Egecioglu, Ö. and Srinivasan, A., 1995. Givens and Householder reductions for linear least squares on a cluster of workstations.

\bibitem{hybrid-petsc}
Lange, M., Gorman, G., Weiland, M., Mitchell, L. and Southern, J., 2013, June. Achieving efficient strong scaling with PETSc using hybrid MPI/OpenMP optimisation. In International Supercomputing Conference (pp. 97-108). Springer, Berlin, Heidelberg.

\end{thebibliography}
\end{document}